\documentstyle[aps,psfig,multicol,amsmath,eqsecnum]{revtex}
\begin{document}
%
%-----------------------------------------------------------------------
\title{Local absorption spectra of artificial atoms and molecules}
%-----------------------------------------------------------------------
\author{C. D. Simserides, U. Hohenester, G. Goldoni, and E. Molinari}
\address{Istituto Nazionale per la Fisica della Materia (INFM) and 
Dipartimento di Fisica, \\
Universit\`a di Modena e Reggio Emilia, Via Campi 213A, 
 I-41100 Modena, Italy}
%-----------------------------------------------------------------------
\maketitle

%%%%%%%%%%%%%%%%%%%%%%
%      ABSTRACT      %
%%%%%%%%%%%%%%%%%%%%%%
\begin{abstract}

We investigate theoretically the spatial dependence of the linear 
absorption spectra of single and coupled semiconductor quantum dots,
where the strong three-dimensional quantum confinement leads to an 
overall enhancement of Coulomb interaction and, in turn, to a pronounced 
renormalization of the excitonic properties. We show that ---because of 
such Coulomb correlations and the spatial interference of the exciton 
wavefunctions--- unexpected spectral features appear whose intensity 
depends on spatial resolution in a highly non-monotonic way 
when the spatial resolution is comparable with the excitonic Bohr radius. 
We finally discuss how the optical near-field properties of double quantum
dots are affected by their coupling.

\end{abstract}

% 78.66.Fd  III-V semiconductors
% 85.30.Vw  Low-dimensional quantum devices (quantum dots, quantum wires
%           etc.)
% 71.35.-y  Excitons and related phenomena
% 73.20.Dx  Electron states in low-dimensional structures (superlattices,
%           quantum well structures and multilayers)

%-----------------------------------------------------------------------
\begin{multicols}{2}
\narrowtext
%

%%%%%%%%%%%%%%%%%%%%%%%%%%%%%%
%        INTRODUCTION        %
%%%%%%%%%%%%%%%%%%%%%%%%%%%%%%
\section{Introduction}

In recent years much attention is being devoted to the properties of 
semiconductor quantum dots (QDs). In these systems, carriers are subject 
to a confining potential in all spatial directions, giving rise to 
a discrete energy spectrum (``artificial atoms'') and to novel phenomena 
of interest for fundamental physics as well as for applications
to electronic and optoelectronic devices \cite{hawrylak,bimberg}. 
The extension and the shape of the QD confining potential varies, 
depending on the nanostructure fabrication technique:
The dots that are studied most extensively by optical 
methods are induced by quantum well (QW) thickness 
fluctuations \cite{zrenner,gammon,hess,flack}, or obtained 
by spontaneous island formation in strained layer epitaxy 
\cite{marzin,grundmann,petroff}, self-organized growth on patterned 
substrates \cite{kapon}, stressor-induced QW potential modulation
\cite{lipsanen}, cleaved edge overgrowth 
\cite{wegscheider}, as well as chemical self-aggregation 
techniques \cite{murray,alivisatos}. The resulting confinement lengths fall 
in a wide range between 1$\mu$m and 10 nm.

In spite of the continuing progress, all the available fabrication
approaches still suffer from the effects of inhomogeneity and 
dispersion in the dot size, which lead to large linewidths when 
optical experiments are performed on large QD ensembles. A major 
advancement in the field has come from different types of local 
optical experiments, that allow the investigation of individual 
quantum dots thus avoiding inhomogeneous broadening
\cite{zrenner,gammon,hess,flack,marzin,grundmann,petroff,kapon,lipsanen,wegscheider,murray,alivisatos}.

Among local spectroscopies, the approaches based on 
scanning near-field optical microscopy (SNOM) \cite{paesler}
are especially interesting as they bring the spatial resolution well below 
the diffraction limit of light: With the development of 
small-aperture optical fiber probes, sub-wavelength resolutions were 
achieved ($\lambda /8-\lambda/5$ \cite{saiki} 
or $\lambda/40$ \cite{betzig}) and the first applications to nanostructures 
became possible \cite{hess,flack,harris,richter,landin,chavez,toda,matsumoto}.
As the resolution increases, local optical techniques in principle 
allow direct access to the space and energy distribution of quantum 
states within the dot. This opens, however, a number of questions 
regarding the interpretation of these experiments, that were often 
neglected in the past.

First of all, for spatially inhomogeneous electromagnetic (EM) fields
it is no longer possible to define and measure an absorption coefficient 
that {\it locally} relates the absorbed power density with the light 
intensity (since the susceptibility $\chi({\bf r,r'})$ cannot be 
approximated by a local tensor). In the linear regime, a local absorption 
coefficient can still be defined, which is however a complicated function 
that depends on the specific EM field distribution 
\cite{mauritz}. The interpretation of near-field spectra therefore 
requires calculations based on a reasonable assumption for the profile 
of the EM field.

Secondly, the quantum states that are actually probed are
few-particle states of the interacting electrons and holes photoexcited 
in the dot. Even in the linear regime, excitonic 
effects are known to dominate the optical spectra of dots since Coulomb 
interactions are strongly enhanced by the three-dimensional 
confinement. Near-field spectra probe exciton 
wavefunctions, and their spatial coherence and 
overlap with the em-field profile will determine the local 
absorption \cite{mauritz}. 

In this paper, we show how the above phenomena affect 
local spectra of QDs, with special attention to the case 
of coupled dots (``artificial molecules'') where carriers 
interact across the barrier via tunneling and/or 
Coulomb matrix elements \cite{rontani2}. Indeed, the optical properties
of coupled dots are currently of great interest not only 
in view of the unavoidable inter-dot interactions occurring 
in real samples with dense QD packing, but also 
in view of their relevance for designing novel devices
including gates for possible solid-state implementations
of quantum information processing \cite{QC}.

We will show that the relative phase of the exciton 
wavefunction in adjacent coupled dots (or in different 
regions of the same dot) can induce dramatic changes in 
the selection rules with respect to far-field spectra:
A realistic prediction of these effects require 
accurate calculations taking into account quantum 
confinement as well as Coulomb interactions. 
Our theoretical scheme is especially designed to allow
a realistic description of the quantum states of the
interacting electron and holes photoexcited in the linear 
regime. In this respect we improve drastically over
previous approaches, which generally focused on a more 
detailed treatment of the EM field distributions
\cite{chang,koch,girard,bryant,chavez-th}.

Our theoretical framework for dots is summarized in 
Section ~\ref{sec:theory}, while Sections~\ref{sec:results} 
and ~\ref{sec:conclusions} discuss our results and conclusions
for single and coupled dots.

%%%%%%%%%%%%%%%%%%%%%%%%
%        THEORY        %
%%%%%%%%%%%%%%%%%%%%%%%%
\section{Theory}
\label{sec:theory}

In this section, we summarize our theoretical approach for computing local
absorption spectra for semiconductor QDs. We first show in
Sec.~\ref{sec:theory.sp} how to compute the single-particle eigenstates for
electrons and holes subjected to a three-dimensional confinement potential.
These single-particle states are then used in 
Sec.~\ref{sec:theory.excitons}
for the calculation of electron-hole (i.e., optical) excitations. In 
analogy
to semiconductor systems of higher dimensionality, we shall refer to these
excitations as {\em excitons};\/ the properties of such excitons, however,
are not only governed by the attractive electron-hole Coulomb interaction
but in addition by the strong quantum confinement. Finally, we use in
Sec.~\ref{sec:theory.local.absorption} the above ingredients to derive the
equations needed for the calculation of local optical absorption spectra.

\subsection{Single-particle states}
\label{sec:theory.sp}

In semiconductor QDs, carriers are confined in all three space
directions.  To simplify our analysis, we assume that a suitable
parameterization of the dot confinement potential is known (e.g., from
experiment) and that the confinement potential varies sufficiently slow
on the length scale of the lattice constant. We thus shall make use of
the envelope-function approach \cite{yu:96}; moreover, since the energy
region of our present concern is relatively close to the semiconductor
band gap, we describe the material band structure in terms of a single
electron and hole band within the usual effective-mass approximation.
More specifically, the envelope-function equation for single electrons
and holes reads:

\begin{equation}\label{eq:sp}
  \bigl\lgroup -\frac{\hbar^2\nabla^2}{2m_{e,h}}
  +V_c^{e,h}({\bf r})\bigr\rgroup
  \phi_\mu^{e,h}({\bf r})=
  \epsilon_\mu^{e,h}\phi_\mu^{e,h}({\bf r}),
\end{equation}

\noindent where $m_e$ ($m_h$) is the effective mass and $V_c^e$
($V_c^h$) is the confinement potential energy for electrons (holes).
Following our approach developed earlier \cite{rossi:96}, we
numerically solve Eq.~(\ref{eq:sp}) for arbitrary confinement
potentials by use of a plane-wave expansion with periodic boundary
conditions (see Appendix A).

\subsection{Exciton states}
\label{sec:theory.excitons}

When the dot structure is perturbed by an external light field (e.g.,
laser), electron-hole pairs are created which propagate in the presence
of the mutual Coulomb interaction and of the dot confinement potential.
Within the present paper, we shall restrict ourselves to the linear
optical response, i.e., the dynamics of a single electron-hole pair.
Then, the exciton dynamics is described by the electron-hole
wavefunction $\Psi({\bf r}_e,{\bf r}_h)$, with the squared modulus
being the probability of finding the electron at position ${\bf r}_e$
when the hole is at position ${\bf r}_h$.

If we expand the electron-hole (``exciton'') eigenfunction in terms of
 single-particle states viz.:

\begin{equation}
  \Psi^{\lambda}({\bf r}_e,{\bf r}_h)=\sum_{\mu\nu}
  \phi_\mu^e({\bf r}_e)\Psi^{\lambda}_{\mu\nu}\phi_\nu^h({\bf r}_h),
\end{equation}

\noindent we obtain the excitonic eigenvalue problem
\cite{haug:93,rossi:98}:

\begin{equation}\label{eq:exciton}
  (\epsilon_\mu^e+\epsilon_\nu^h)\Psi_{\mu\nu}^\lambda+
  \sum_{\mu'\nu'}V_{\mu\mu',\nu\nu'}^{eh}\Psi_{\mu'\nu'}^\lambda=
  E_\lambda \Psi_{\mu\nu}^\lambda.
\end{equation}

\noindent As will be shown in the following, the exciton spectrum
$E_\lambda$ directly provides the optical transition energies whereas
the excitonic wavefunctions $\Psi^\lambda$ determine the oscillator
strengths of the corresponding transitions. In Eq.~(\ref{eq:exciton}) we
have introduced the electron-hole Coulomb matrix elements
\cite{remark:coulomb.eh}:

\begin{equation}\label{eq:coulomb.eh}
  V_{\mu\mu',\nu\nu'}^{eh}=-e^2\int d{\bf r}_ed{\bf r}_h\;
  \frac{{\phi_\mu^e}^*({\bf r}_e)\phi_{\mu'}^e({\bf r}_e)
        {\phi_\nu^h}^*({\bf r}_h)\phi_{\nu'}^h({\bf r}_h)}
       {\kappa_o|{\bf r}_e-{\bf r}_h|},
\end{equation}

\noindent where $e$ is the elementary charge and $\kappa_o$ is the
static dielectric constant of the bulk semiconductor (note that in
Eq.~(\ref{eq:coulomb.eh}) we have not considered the electron-hole
exchange interaction). Within our computational approach, we consider
in Eq.~(\ref{eq:exciton}) typically a basis of 12 states for electrons
and holes, respectively, and obtain the excitonic eigenfunctions by
direct diagonalization of the Hamiltonian matrix.

\subsection{Local optical absorption}
\label{sec:theory.local.absorption}

When the semiconductor nanostructure is excited by a local near-field
probe, the total absorbed power $\alpha(\omega)$ at a given frequency
$\omega$ is proportional to $\int d{\bf r}\;{\cal E}_\omega ({\bf
r})P({\bf r},\omega)$, where ${\cal E}_\omega({\bf r})$ is the 
electro-magnetic field distribution of the near-field probe.
Within linear response, the induced interband
polarization $P({\bf r},\omega)$ is related to ${\cal E}_\omega({\bf
r})$ through:

\begin{equation}
  P({\bf r},\omega)=\int d{\bf r}'\; 
  \chi({\bf r},{\bf r}';\omega){\cal E}_\omega({\bf r}'),
\end{equation}

\noindent where the non-local electrical susceptibility $\chi({\bf
r},{\bf r}';\omega)$ can be expressed in terms of the excitonic
eigenenergies and eigenfunctions \cite{mauritz}:

\begin{equation}
  \chi({\bf r},{\bf r}';\omega)=\mu_o^2\sum_\lambda
  \frac{\Psi^\lambda({\bf r},{\bf r}) {\Psi^\lambda}^*({\bf r}',{\bf r}')}
  {E_\lambda-\hbar\omega-i\gamma}.
\end{equation}

\noindent Here, $\mu_o$ is the dipole-matrix element of the bulk 
semiconductor, while we have introduced a small damping constant $\gamma$
accounting for the finite lifetime of exciton states due to environment
coupling (e.g., phonons). To derive our final expression, it turns
out to be convenient to consider for the elctromagnetic field
distribution a given profile $\xi$ centered around the beam position
${\bf R}$, i.e., ${\cal E}_\omega({\bf r})={\cal E}_\omega\xi({\bf
r}-{\bf R})$. Then, the local spectrum for a given tip position ${\bf
R}$ can be expressed in the form \cite{mauritz} (see also Appendix A):

\begin{equation}\label{eq:local.spectrum}
  \alpha_\xi({\bf R},\omega) \propto 
  \Im   \sum_{\lambda} 
  \frac {\alpha_\xi^\lambda ({\bf R}) } {E_\lambda-i\gamma-\hbar\omega}, 
\end{equation}

\noindent where 

\begin{equation}\label{eq:lambda}
  \alpha_\xi^\lambda ({\bf R}) = \left|\int d{\bf r} \;
  \Psi^{\lambda}({\bf r},{\bf r}) \xi({\bf r-R}) \right|^2.
\end{equation}

\noindent Two limiting cases can be identified.
For a spatially homogeneous electromagnetic
field (far-field), the oscillator strength $\alpha_\xi^\lambda$ is given by 
the
spatial average of the excitonic wavefunction, i.e.,
$\alpha_\xi^\lambda ({\bf R})=|\int d{\bf r} \; \Psi^{\lambda}({\bf
r},{\bf r})|^2$. In the opposite (and hypothetical) limit of an
infinitely narrow probe, $\xi({\bf r-R})=\delta({\bf r-R})$, one is
probing the local value of the exciton wavefunction, i.e.,
$\alpha_\xi^\lambda ({\bf R}) = |\Psi^{\lambda}({\bf R},{\bf R})|^2$.
Finally, within the intermediate regime of a narrow but finite probe,
$\Psi^{\lambda}({\bf r},{\bf r})$ is averaged over a region which is
determined by the spatial extension of the light beam; therefore,
excitonic transitions which are optically forbidden in the far-field
may become visible in the near-field.

%%%%%%%%%%%%%%%%%%%%%%%%%%%%%%%%%%%%%%%%
%        SINGLE AND DOUBLE DOTS        %
%%%%%%%%%%%%%%%%%%%%%%%%%%%%%%%%%%%%%%%%
\section{Results}
\label{sec:results}

In the following sections we consider the interaction 
of the EM field with excitonic states of single and double QDs; for the
latter system, we focus particularly on the transition between two isolated
QDs and a the ``artificial molecule'', where the electronic states
of two QDs are strongly overlapping.

\subsection{Single-particle states}

We shall consider a prototypical QD confinement which is
composed of a 2D harmonic potential in the $(x,y)$-plane and a
rectangular quantum well along $z$; such confinement potentials have
been demonstrated to be a good approximation for
self-assembled quantum dots formed by strained-layer epitaxy. 
We focus on cases where the $z$-confinement is stronger
than the $(x,y)$ one, so that the confinement potential can be written
as

\begin{equation}\label{eq:total.confinement}
  V_c^{e,h}(x,y,z)=V_\|^{e,h}(x,y)+
                   V_o^{e,h}\theta(|z|-\frac{z_o}2),
\end{equation}

\noindent where $z_o$ is the width of the quantum-well and $V_o^{e,h}$ the
band offsets for electrons and holes, respectively. For a single dot
the in-plane confinement potential $V_\|^{e,h}(x,y)$ is of the form:

\begin{equation}\label{eq:confinement.single}
  V_\|^{e,h}(x,y)=\frac 1 2 {\cal K}_{e,h}(x^2+y^2),
\end{equation}

\noindent while for two dots (i.e., double dot) separated by the distance
$d$:

\begin{equation}\label{eq:confinement.double}
  V_\|^{e,h}(x,y)=
  \begin{cases}
      \frac 1 2{\cal K}_{e,h}\bigl((|x|-\frac d 2)^2+y^2\bigr) & 
      \mbox{for $|x|>\frac d 4$} \\
      \frac 1 2 {\cal K}_{e,h}\bigl((\frac {d^2} 8 - x^2)+y^2\bigr) &
      \mbox{otherwise}\\
  \end{cases}
\end{equation}

\noindent with ${\cal K}_{e,h}=m_{e,h}(\omega_o^{e,h})^2$, and
$\hbar\omega_o^{e,h}$ the level splittings of the in-plane harmonic
potential. The shape of the double-dot potential has been obtained by
matching the parabolas with opposite curvature, such that the potential
is continuous and smooth at $x=\pm\frac d 4$; the shape of the
resulting potential along the $x$- direction is shown, for selected
inter-dot distances $d$, in Figs. 1(b), 1(d). Material and dot
parameters which are used in this paper are listed in Table
\ref{table:dot}; with this choice of parameters, electron and hole
wavefunctions have approximately the same lateral extension, and the
QW-induced intersubband splittings are much larger than
$\hbar\omega_o^{e}$ and $\hbar\omega_o^{h}$.

With our choice of the confinement potential,
Eq.~(ref{eq:total.confinement}), the single-particle energies of a QD
are $E_{\mbox{\tiny QD}} = E_{\mbox{\tiny QW}} + E_{\mbox{\tiny
harm}}$, where $E_{\mbox{\tiny QW}} $ is the confinement energy of the
QW along $z$ and $E_{\mbox{\tiny harm}}$ is the confienment energy of
the 2D single- or double-harmonic potential. Single-particle energies
and envelope functions have been computed numerically within a
plane-wave scheme. However, for a single QD the 2D eigenstates can be
found analytically and are the well-known ``Fock-Darwin'' states
\cite{hawrylak} (we stress, however, that the extension in the
$z$-direction is of crucial importance for the calculation of the
Coulomb matrix elements and the optical properties, and unavoidably has
to be taken into account in any realistic calculation; see also
discussion in Ref.~\cite{rontani}). For such states $E_{\mbox{\tiny
harm}}= (n+1) \hbar\omega_o^{e,h}$, where $n=0,1,\ldots$ is the
principal quantum number, and each level is $(n+1)$-fold degenerate;
in Table~\ref{table:fock.darwin} we summarize for convenience 
some properties of these ``Fock-Darwin'' states \cite{hawrylak}.

Figure 1 shows the calculated single-particle energies 
for electrons and holes for the more complex case of 
a double QD with the confinement
potential given in Eq.~(\ref{eq:confinement.double}) and with 
parameters listed in Table \ref{table:dot}. The lower
panels (Figs.~1(b,d)) show the confinement potentials for electrons and
holes at selected inter-dot distances. 
Obviously, for large dot separations $d$ ($d\gtrsim 60$ nm)
the system can be well approximated by two separate
QDs; in this regime the equidistance of the excited states and the 
correct degeneracy of the Fock-Darwin states is obtained.
When $d$ is small enough that carriers 
have sufficient energy to overcome (or tunnel through) the barrier
between the two dots, the degeneracy is removed, and the energy levels
have a non-monotonous behaviour which reflects the
transition from two separated carrier systems to a single one,
and is similar to the one found, e.g. for coupled QWs \cite{costas}.
For the smallest dot distances the double-dot potential merges into a
single-dot potential, and the Fock-Darwin states of a single dot are 
recovered. 

\subsection{Role of the Coulomb correlation in the far-field spectra}

Before turning to the analysis of near-field spectra, we shortly
discuss the limiting case of very broad EM field distribution 
(far-field spectra). This discussion allows us
to elucidate the role of the electron-hole Coulomb correlation, 
particularly in the transition from two separate ``artificial atoms''
to an ``artificial molecule''.

Far-field spectra  can be obtained in the formalism of 
Sec.\ \ref{sec:theory.local.absorption}
with  a spatially homogeneous electromagnetic field distribution
probe $\xi$):

\begin{equation}\label{eq:far.field}
  \alpha_{\xi({\bf r})={\rm const}}^\lambda  = \left|\int d{\bf r} \;
  \Psi^{\lambda}({\bf r},{\bf r}) \right|^2.
\end{equation}

Figs.~2 and 3 show the calculated far-field spectra for a double QD as
a function of the dot distance $d$. We first concentrate on the
calculations where Coulomb correlations were artificially set to zero
(Figs.~2(a) and 3(a)):
Because of symmetry, only a
small fraction of all possible electron-hole transitions is visible; 
from Eq.~(\ref{eq:far.field}) and using $\int d\varphi \exp
i(m_e+m_h)\varphi\propto\delta_{m_e,-m_h}$ we obtain that optical
transitions are only allowed between electron and hole single-particle
states with opposite angular momentum. Indeed, 
for large distances
(uncoupled QDs) only three strong absorption peaks are observed, with an
energy splitting of approximately $\hbar\omega_o^e+\hbar\omega_o^h$;
the intensity of the peaks increases with energy (with ratio $1:2:3$).
These can be attributed to transitions between electrons
and holes  single-particle states (see Table
\ref{table:fock.darwin}) of the $1s$ symmetry  (peak at $\approx 70$ meV), 
the $1p$ symmetry (peak at $\approx
95$ meV), and the $1d$ and $2s$ symmetries (peak at $\approx 120$ meV).

When symmetry is reduced, either because of an asymmetric confinement
potential or by the presence of an external inhomogeneous EM-field 
(as will be 
discussed later), the selection rules noted above are relaxed. Indeed,
when $d$ is reduced and the two QDs begin to interact, the calculated
spectra show a much richer structure, as shown in Fig.~2(a) and 3(a),
reflecting the reduction of built-in symmetry. Obviously, when
$d\simeq 0$, the usual selection rules of a single-QD are recovered.

When Coulomb interaction is included, inspection of the exciton
wavefunctions $\Psi_{\mu\nu}^\lambda$ (obtained from the solutions of
Eq.~(\ref{eq:exciton})) shows that a number of different
single-particle transitions contributes to each excitonic state 
\cite{basis}.
Coulomb interaction affects the optical spectra [see
Figs.~3(a) and 3(b)] in several ways. Firstly, because of the attractive
electron-hole interaction leading to the ``bound'' excitonic states,
we observe a red-shift of the peaks; peak separation is barely
affected, however, at least for large $d$. Secondly, we observe a
redistribution of the oscillator strength; for isolated QDs the three
main peaks are of similar height. In general, oscillator strength is
transferred from higher to lower peaks. This effect is particularly
strong, e.g., in the doublet which splits from the lowest peak when
the two QDs approach; contrary to the uncorrelated case, the heighest
partner is very weak.  Finally, Coulomb interaction is responsible for
the appearance of additional lines (see, e.g., for $d=70$ nm the peak
at 70 meV). While the first two effects (red-shift and transfer of
oscillator strength) are similar to what is found in the absorption
spectra of semiconductor quantum wires, and thus can be considered as
a general fingerprint of Coulomb correlations in the optical
properties of semiconductor nanostructures, the origin of the additional 
peaks is best discussed in connection with the calculated near-field
optical spectra and, therefore, is postponed to the next section.

\subsection{Optical near-field spectra}

In this section we discuss the local absorption spectra of single and
coupled QDs. Because of the narrow well width of the dot confinement
potential (see Table \ref{table:dot}), the EM profile of
the near-field probe along $z$ has an only minor influence on the
results, and we use:

\begin{equation}\label{eq:em.profile}
  \xi(x,y,z)\propto\exp\bigl\lgroup-\frac{x^2+y^2}{2\sigma^2}
  \bigr\rgroup.
\end{equation}

\noindent The spatial resolution of the electromagnetic field
distribution of Eq.~(\ref{eq:em.profile}) is then approximately given
by the full-width at half maximum (FWHM) of the Gaussian (i.e., $2
\sqrt{2 \; \ln 2} \sigma \approx 2.35 \sigma$). Since the Gaussian acts
as an envelope on $\Psi^{\lambda}$, in the intermediate regime of a
narrow but finite $\sigma$ the spatial average only extends over the
region where the Gaussian is non-vanishing.

Since the extension of the quantum states under investigation is of
the order of a few tens of nano-meters (see also Figs.~5 and 8, to be
discussed below), in our calculations we consider three different
regimes of spatial resolution: {\em i)} a regime where the FWHM is
much larger than the extension of the quantum states (as a
characteristic value we use $\sigma=50$ nm); {\em ii)} a regime where
the FWHM is comparable to the extension of the relevant quantum states
(we use $\sigma=10$ nm); {\em iii)} a regime with an
extremely narrow probe beam (we use $\sigma=0.1$ nm). Calculations 
performed in this latter (unphysical) regime are used for illustrative 
purposes to obtain a ``cartography'' of the exciton wavefunction, 
as discussed at the end of Sec.\ \ref{sec:theory.local.absorption}. 
We finally notice that the excitonic Bohr radius $\approx 12$ nm for GaAs.

\subsubsection{Single quantum dot}

In Fig. 4 we report the calculated local absorption spectra
$\alpha_\xi(X,\hbar
\omega)$ for a single QD as a function of the tip position. The tip
is swept along one direction, passing through the center of the QD. 

In Figs.~4(a--c) we show the calculated spectra neglecting Coulomb
interaction. For the
highest spatial resolution [Fig.~4(a)], the local absorption at photon
energy $E_\lambda$ is proportional to $\int dz\;\Psi^\lambda({\bf
r},{\bf r})|_{y=0}$. Given the energy splitting $\hbar\omega_o^h=3.5$
meV for holes and $\hbar\omega_o^e=20$ meV for electrons, we can
attribute the triplet of peaks at $\approx 70$ meV to the
single-particle transitions involving the $1s$ state of electrons and
the $1s$, $1p$, and $(2s,1d)$ states (in order of increasing energy)
of holes (see also Table \ref{table:fock.darwin}); analogously, the
triplet at $\approx 90$ meV is attributed to the transitions involving
the $1p$ state of electrons and the $1s$, $1p$, and $(2s,1d)$ hole
states; indeed, in Fig.~4(a) the localization of the absorption peaks
is suggestive of the $s$-, $p$- or $d$-type symmetry of the
corresponding Fock-Darwin states. These features are still present
at the intermediate resolution [Fig.\ 4(b)], 
but disappear at the opposite limit of a broad
probe [Figs.~4(c)]. This is expected, since, when a localized EM-field
is present, the symmetry of the whole system (nanostructure+EM-field) is
lower than that of the nanostructure (except when the probe is centered
in the symmetry center of the structure), and far-field selection rules are
relaxed. When the probe is broadened, however, 
the built-in symmetry of the structure is recovered, and 
optical far-field selection rules (i.e.,
optical transitions only between electron and hole states with
opposite angular momentum $m$) apply; therefore, the spectra are
almost identical to those of two separated dots in far-field
spectroscopy, already discussed in Figs.~2 and 3.

When we compare Figs.~4(a--c) with Figs.\ 4(d--f), we find that Coulomb
interaction induces several effects which are expected on the basis of
the discussion of the far-field spectra. In particular, we find {\em
i)} an almost rigid redshift of the spectra; {\em ii)} a transfer of
oscillator strength from transitions at higher energies to those at
lower energies; {\em iii)} the appearance of new features in the
optical spectra. To discuss the origin of these new optical features
caused by Coulomb interactions, let us consider, e.g., the optical
peaks at photon energy $\approx 65$  meV (Figs.~4(d--f)): They are
quite strong at $\sigma=0.1$ nm (Fig.~4(d)), almost disappear at
$\sigma=10$ nm (Fig.~4(e)), and are visible again in the far field
limit (Fig.~4(f)). Such a behaviour is rather unexpected and noticeably
differs from that of other transitions, which ---with increasing
$\sigma$--- either remain strong or gradually disappear due to
symmetry reasons, as discussed above. To investigate the origin of this
non-monotonic dependence, in the following we analyze the three
excitons within the corresponding energy range. Figure 5 shows a
contour plot of the respective exciton wavefunction $\Psi^\lambda({\bf
r},{\bf r})|_{z=0}$. Apparently, in Fig.~5(a) the exciton has $s$-type
symmetry, whereas the other two electron-hole states have $p$-type
symmetry. (Because of the periodicity box used in our calculations, the
two-fold degenerate $p$-type exciton wavefunctions have cartesian
rather than cylinder symmetry; note that, since the presence of the
near-field tip destroys the cylinder symmetry, the wavefunctions shown
in Fig.~5 indeed form a natural basis; see also Table
\ref{table:fock.darwin}). Next, we note that the average $\int d{\bf
r}\;\Psi^\lambda({\bf r},{\bf r})$ of the $p$-type exciton
wavefunctions is zero. Since with increasing $\sigma$ the radius within
which the exciton eigenfunctions $\Psi^{\lambda}$ are averaged
increases, we expect for these $p$-type functions with increasing
$\sigma$ a monotonically decreasing behavior. The exciton shown in
Fig.~5(a), on the other hand, has a non-zero average and is therefore
visible in both the optical far- and near-field. A closer inspection of
the exciton wavefunction $\Psi_{\mu\nu}^\lambda$ reveals that the
largest contribution stems from the transition between the $1s$ state
of electrons and the $2s$ state of holes, but there is also a
noticeable contribution from the $1s$-$1s$ and $1p$-$1p$ electron-hole
transitions. Indeed, only the latter contributions couple in the
far-field to the light field. In the regime of finite resolution, there
is an optimal cancellation when the FWHM of the EM near field 
(i.e. $2 \sqrt{2 \; \ln 2} \sigma \approx 2.35 \sigma$)
becomes equal to the Bohr radius. This is clearly depicted in Fig.~6,
where, in order to  facilitate our discussion, we have introduced the 
quantity $I_\xi^\lambda\propto\int d{\bf R}\;\alpha_\xi^\lambda ({\bf R})$,
which provides a measure of the relative contribution of each exciton
to the absorption spectra. Fig.~ 6 shows $I_\xi^\lambda$ for the three
excitons (shown Fig.~5) within the energy region of 65 meV: 
We observe that with increasing $\sigma$, 
the $p$-type functions (open circles) indeed vanish monotonically, 
whereas for the $s$-type exciton (full circles) 
there exists an optimal cancellation when the FWHM of the EM
field distribution becomes approximately equal to the Bohr radius. 
In spite of the specific carrier states of a single parabolic QD, 
we expect that such non-monotonic behavior appears quite generally in 
semiconductor nanostructures where carrier states are confined on a 
length scale comparable to the Bohr radius, and thus provides a striking 
fingerprint of Coulomb correlations in the optical near-field spectra 
(we find similar behavior in our calculations for the near-field spectra 
of coupled QDs discussed below).

\subsubsection{Double quantum dot}

In Figure 7 we show the calculated local absorption spectra 
$\alpha_\xi(X,\hbar\omega)$ for a double QD for selected values of the 
interdot distance and $\sigma$. The tip of the probe is swept along the 
direction which passes through the centers of the two QDs.  

Let us first concentrate on the results with $\sigma=0.1$ nm and with 
the Coulomb interaction taken into account (Figs.~7(d,g,j)). 
With decreasing interdot distance we observe the transition from a system 
where the energetically lowest exciton states are almost localized in the 
spatially separated minima of the two dots, to a system where the 
electron-hole states extend over the whole nanostructure.
 Here, the $s$-like ground-state excitons of Fig.~7(j)
split up into a ``bonding'' and an ``anti-bonding'' state (Fig.~7(d)).
By comparing Figs.~7(d) and 7(f), we find that in the optical far-field
only the symmetric ground state exciton couples to the light field.

Next, we discuss the optical features at the photon energy of $\approx 70$ 
meV for $d=40$ nm. As in the case of the single dot, these features show a
non-monotonic dependence on the probe width. As can be inferred from
the calculations with $\sigma=0.1$ nm, there are several excitonic
states contributing to the spectral features in this energy range;
Fig.~8 shows the excitonic wavefunction of two states out of the six
states with $E_\lambda\approx 70$ meV for for $d=40$ nm; it can be
inferred that for a spatial resolution of the near-field probe
comparable to the excitonic Bohr radius ($\approx 12$ nm) there is
again an optimal cancellation. This is a remarkable
finding, because it clearly demonstrates that such a behaviour indeed
is a general characteristics of semiconductor nanostructures, and does
not depend on peculiar symmetries of the confining potential.

%%%%%%%%%%%%%%%%%%%%%%%%%%%%%%%%%%%%%%%%%
%        SUMMARY AND CONCLUSIONS        %
%%%%%%%%%%%%%%%%%%%%%%%%%%%%%%%%%%%%%%%%%
\section{Summary and Conclusions}
\label{sec:conclusions}

We have analyzed theoretically the interaction between a model near-field 
probe and a zero-dimensional heterostructure: Quantum confinement of the 
electron and hole states, as well as their Coulomb interaction in the 
linear regime are fully included in our description. 

We have specifically considered single and coupled semiconductor quantum
dots, and shown that absorption is strongly influenced by the spatial 
interference of the exciton wavefunctions, which depends on the 
spatial extension of the light beam.  As a consequence, near-field 
experiments on quantum dots are predicted to display unexpected spectral 
features whose dependence on spatial resolution is highly non-trivial.

When combined with an appropriate choice of the EM field distribution, 
our approach provides the necessary tool for interpretation of near-field 
absorption spectra of quantum dots as the spatial resolution of experiments 
becomes comparable with the Bohr radius of the exciton in the 
nanostructure.

\acknowledgments

We thank Fausto Rossi for most stimulating discussions. This work was
supported in part by INFM through PRA-99-SSQI, and by the EC under 
the TMR Network ``Ultrafast Quantum Optoelectronics'' and the IST 
programme ``SQID''. U.H. acknowledges support by the EC through a TMR
Marie Curie Grant.

\begin{appendix}

\section{Plane-wave approach}

In this Appendix we discuss details of our numerical solution schemes
based on a plane-wave expansion.  Following our approach developed
earlier \cite{rossi:96}, we consider the problem of a single or double
QD which is located inside a box with periodic boundary conditions,
where the boxsize is chosen sufficiently large to avoid interactions
with ``neighbour'' dots. As a complete set of functions, inside the
periodicity box we use a plane-wave basis, $|{\bf k}\rangle$, with:

\begin{equation}
  k_{\alpha} = \frac {2 \pi n_{\alpha}}{L_{\alpha}}, \;\;
  n_{\alpha} \in {\mathcal{Z}}, \;\; \alpha=x,y,z. 
\end{equation}

\noindent Here $L_\alpha$ denotes the sizes of the periodicity box (we
use the same box for electrons and holes). We next expand the
single-particle wavefunctions for electrons and holes within the
plane-wave basis:

\begin{equation}
  \tilde\phi_{\mu,{\bf k}}^{e,h}=\Omega^{-1}\int d{\bf r}\;
  e^{-i{\bf k}\cdot{\bf r}}\phi_\mu^{e,h}({\bf r}),
\end{equation}

\noindent with $\Omega$ the volume of the periodicity box. The
envelope-function equation (\ref{eq:sp}) is then transformed to:

\begin{equation}
  \sum_{{\bf k}'}\bigl\lgroup\frac{\hbar^2{\bf k}^2}{2m_{e,h}}
  \delta_{{\bf kk}'}+
  \tilde V_{c,{\bf k}-{\bf k}'}^{e,h}\bigr\rgroup
  \tilde\phi_{\mu,{\bf k}'}^{e,h}=
  \epsilon_\mu^{e,h}\tilde\phi_{\mu,{\bf k}}^{e,h},
\end{equation}

\noindent which can be solved by standard diagonalization techniques.
To keep the numerics tractable, only wavevectors smaller than a given
cut-off wavevector are considered (typically 2000--3000 wavevectors).
In our computational approach, we perform the Fourier transform of the
confinement potential by storing $V_c^{e,h}({\bf r})$ on an appropriate
grid (with a typical number of 30 points along each direction), and
approximating within each cube $V_c^{e,h}({\bf r})$ by its average
value.

In the calculation of the near-field spectra, we define the
electron-hole index $l=(\mu,\nu)$. Then:

\begin{equation}
  \Psi^{\lambda}({\bf r},{\bf r}) = \sum_l \Psi_l^{\lambda}\;
  \phi^e_{\mu_l}({\bf r}) \phi^h_{\nu_l}({\bf r}),
\end{equation}

\noindent and we obtain for $\alpha_\xi^\lambda ({\bf R})$ of
Eq.~(\ref{eq:lambda}) the final result:

\begin{equation}
  \alpha_\xi^\lambda ({\bf R})=
  | \sum_l \Psi_l^{\lambda} \sum_{{\bf k},{\bf k}'} 
  \tilde\xi_{{\bf k}+{\bf k}'}({\bf R})\;
  \tilde\phi_{\mu_l,{\bf k}}^{e}\tilde\phi_{\nu_l,{\bf k}'}^{h}|^2,
\end{equation}

\noindent with $\tilde\xi_{\bf k}({\bf R}) = {\Omega}^{-1}
\int d{\bf r} \; \xi({\bf r}) {e}^{i {\bf k} \cdot ({\bf
r}+{\bf R}) }$.

\end{appendix}

\begin{table}
\caption{
Material parameters for GaAs/AlGaAs and dot parameters which were 
used in the calculations. Here, $m_o$ is the free-electron mass.
}

\begin{tabular}{lll}\label{table:dot}
description & value & units  \\ \tableline
electron mass $m_e$  & $0.067$ & $m_o$ \\
hole mass $m_h$      & $0.38$  & $m_o$ \\
dielectric constant $\kappa_o$ & $12.9$ &  \\
conduction-band offset for electrons $V_o^e$ & 300 & meV \\
valence-band offset for holes $V_o^h$ & 200 & meV \\
confinement energy $\hbar\omega_o^e$ for electrons & $20$ & meV  \\
confinement energy $\hbar\omega_o^h$ for holes & $3.5$ & meV \\
quantum-well width $z_o$ & $10$ & nm \\
\end{tabular}
\end{table}

\end{multicols}
\widetext

\begin{table}
\caption{
Eigenfunctions (Fock-Darwin states) with lowest energies for a particle
with mass $\mu$ and for a potential of the form 
$V(x,y)=\frac 1 2 \mu\omega_o^2(x^2+y^2)=\frac 1 2\mu\omega_o^2 r^2$ 
(i.e., two-dimensional harmonic oscillator). We use ${\cal X}=x/a_o$, 
${\cal Y}=y/a_o$, and ${\cal R}=r/a_o$, 
with $a_o = \sqrt(\frac{\hbar}{\mu \omega_o})$.
Because of cylindrical symmetry, the angular momentum in the $z$-direction
is a good quantum number ($m$) and the angular part of the
wavefunctions is of the form $\propto\exp\pm im\varphi$; we use the
notation $s$ for $m=0$, $p$ for $m=\pm 1$, and $d$ for 
$m=\pm 2$.
}

\begin{tabular}{c|c|cc}\label{table:fock.darwin}
                                                &
cartesian coordinates:                          &
\multicolumn{2}{c}{cylinder  coordinates:}      \\
Energy ($\hbar\omega_o$)                        &
$\phi({\cal X},{\cal Y})\propto
 \exp-\frac 1 2({\cal X}^2+{\cal Y}^2)\qquad$   &
$\phi({\cal R},\varphi)\propto
 \exp-\frac 1 2{\cal R}^2\qquad$                &
notation                                        \\
\tableline\tableline
1 & $\times 1$ & $\times 1$ & $1s$ \\ \tableline
$2$ & $\times{\cal X}$ & & \\
& $\times{\cal Y}$ & $\times{\cal R}\exp\pm i\varphi$ & $1p$ \\ \tableline
3 & $\times{\cal XY}$ & & \\
& $\times(2 {\cal X}^2-1)$ & $\times({\cal R}^2-1)$ & $2s$ \\
& $\times(2 {\cal Y}^2-1)$ & $\times{\cal R}^2\exp\pm 2i\varphi$ & $1d$\\
\end{tabular}
\end{table}

%\begin{multicols}

%-----------------------------------------------------------------------
%%%%%%%%%%%%%%%
%   FIGURES   %
%%%%%%%%%%%%%%%

\begin{figure}
\vspace*{0.0cm}
\hspace*{0.4cm}\psfig{file=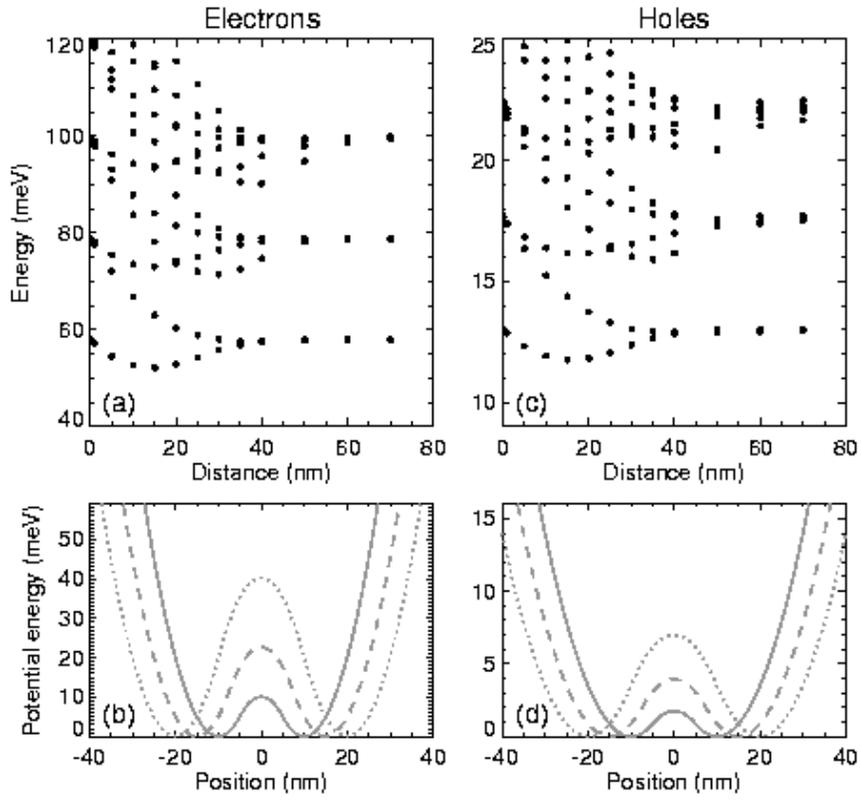,height=20cm}
\vspace*{0.0cm}
\caption
{Single-particle confinement energies as a function of the distance
between the two dots, $d$ (upper panels) and the form of the confining
potential along x-axis (lower panels) for $d = 20$ nm (solid line), $d
= 30$ nm (dashed line) and $d = 40$ nm (dotted line). Left and right
panels correspond to electrons and holes, respectively.}
\end{figure}
\pagebreak

\begin{figure}
\vspace*{0.0cm}
\hspace*{0.4cm}\psfig{file=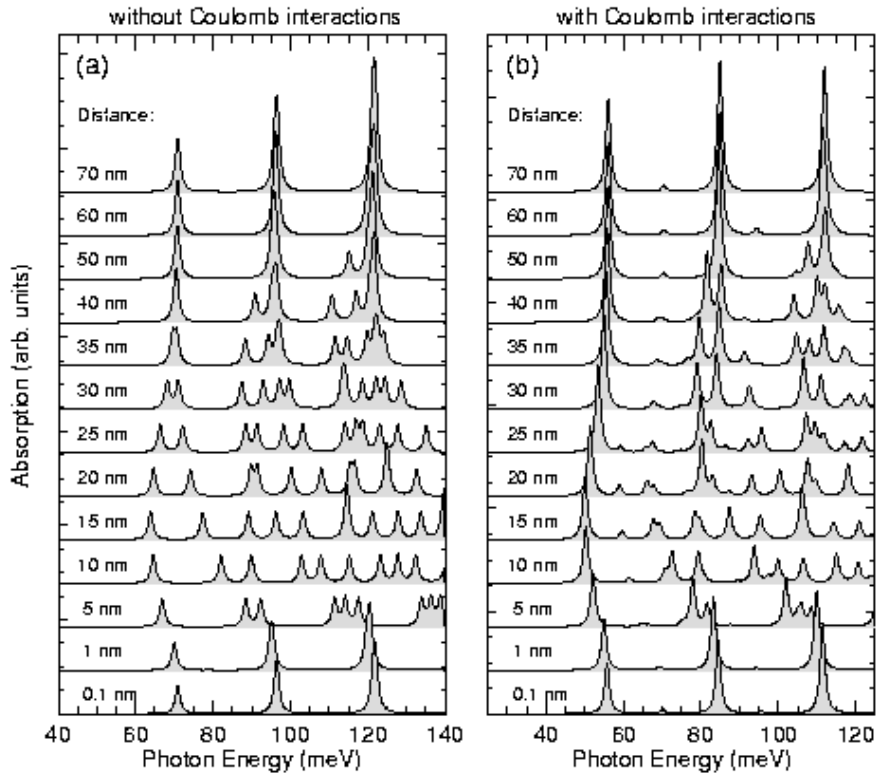,height=20cm}
\vspace*{1.0cm}
\caption
{Optical absorption spectra for a homogeneous electromagnetic field
profile (i.e., far field) for a double quantum dot and for different
distances $d$: (a) Coulomb interactions neglected; (b) Coulomb
interactions included. We use $\gamma=1$ meV. The photon energy is
measured with respect to the bandgap.}
\end{figure}
\pagebreak

\begin{figure}
\vspace*{0.0cm}
\hspace*{0.4cm}\psfig{file=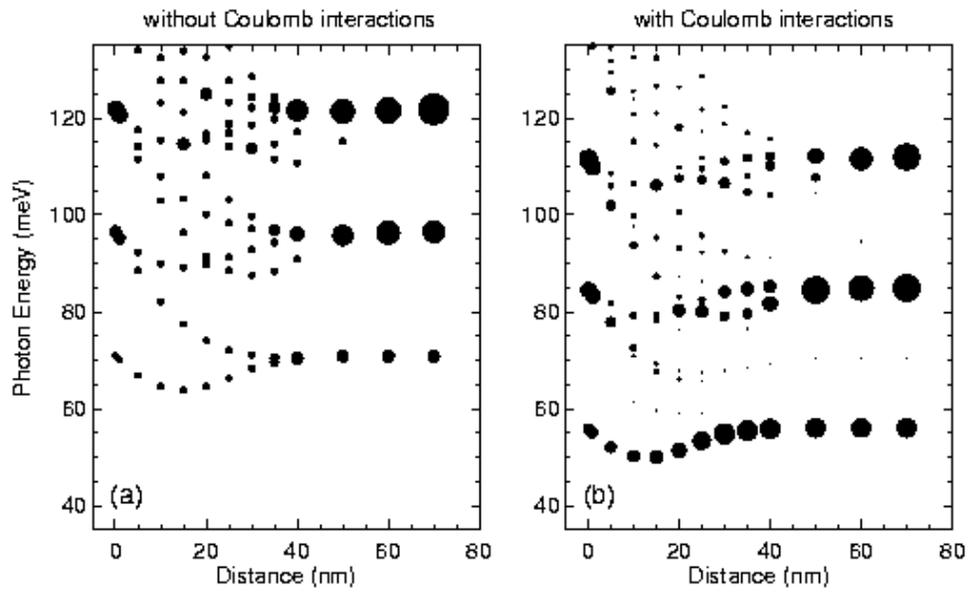,height=20cm}
\vspace*{1.0cm}
\caption
{Same as Fig.~2; the size of each dot corresponds to the height (i.e.,
oscillator strength) of the corresponding absorption peak.}
\end{figure}
\pagebreak

\begin{figure}
\vspace*{0.0cm}
\hspace*{0.4cm}\psfig{file=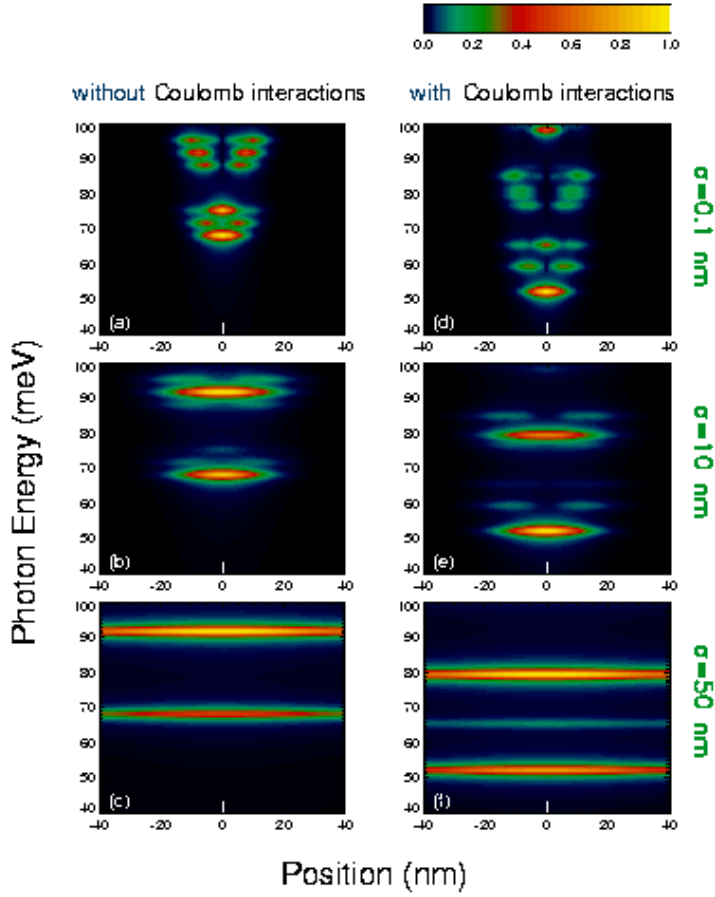,height=20cm}
\vspace*{0.2cm}
\caption
{Local absorption spectra $\alpha_\xi(X,\hbar \omega)$  for a single QD
with (Figs.~4(d--f)) and without (Figs.~4(a--c)) Coulomb interactions
and for different values of $\sigma$. Photon energy $\hbar\omega$ is
measured with respect to the bandgap, and $X$ is the position of the
tip along the $x$-axis ($Y=0$). In this calculations we use a basis of 6
electron and hole states, respectively.}
\end{figure}
\pagebreak

\begin{figure}
\vspace*{0.0cm}
\hspace*{0.4cm}\psfig{file=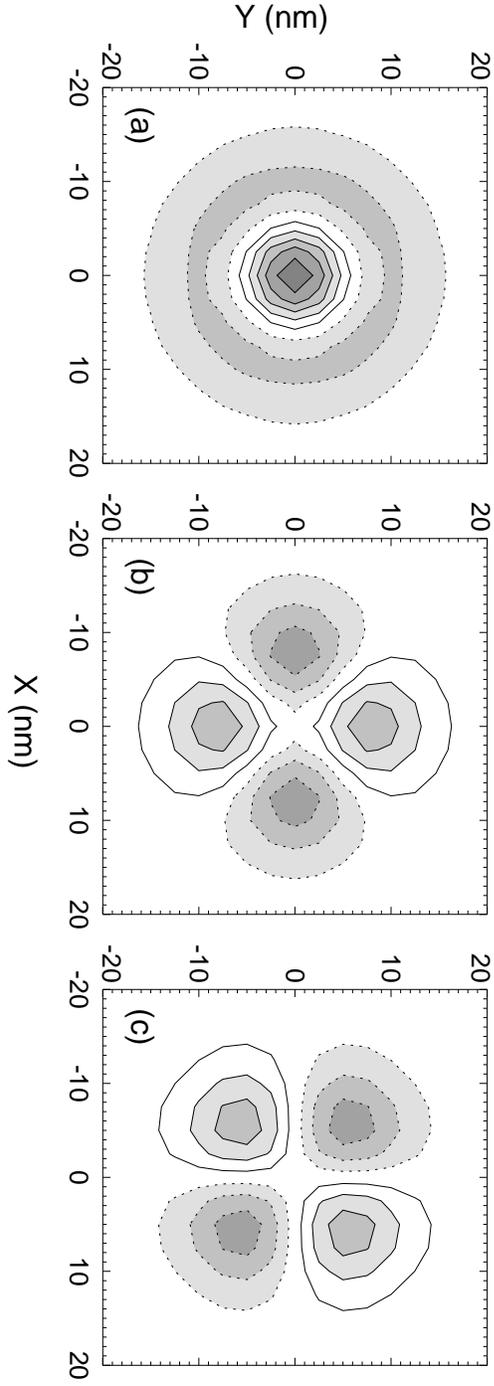,height=20cm}
\vspace*{0.0cm}
\caption
{Contour plot of the exciton wavefunction $\Psi^\lambda({\bf
r},{\bf r})$ for three excitons which contribute to the absorption peak
at $\approx 65$ meV. Solid and dashed lines correspond to positive and
negative values, respectively.}
\end{figure}
\pagebreak

\begin{figure}
\vspace*{0.0cm}
\hspace*{0.4cm}\psfig{file=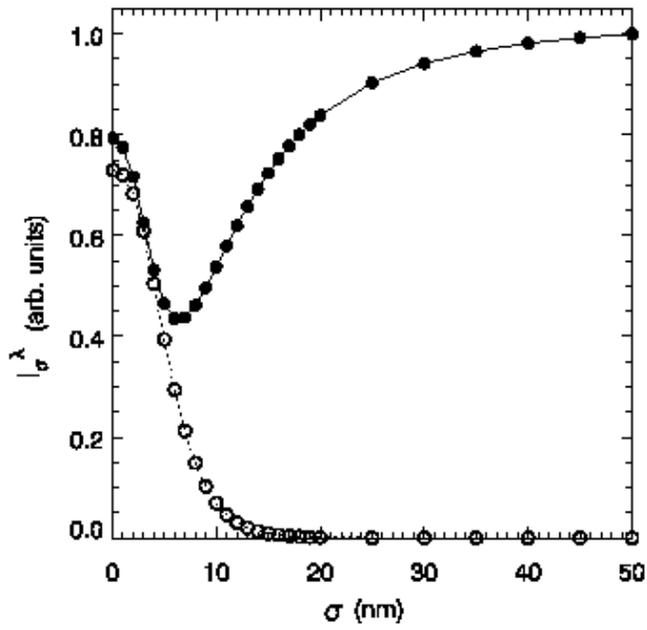,height=20cm}
\vspace*{0.2cm}
\caption
{The relative contribution, $I_\xi^\lambda$, as a function of
$\sigma$, for the excitons (depicted in Fig.5) which are responsible 
for the non-monotonic behavior of the feature at 65 meV.}
\end{figure}
\pagebreak

\begin{figure}
\vspace*{0.0cm}
\hspace*{0.4cm}\psfig{file=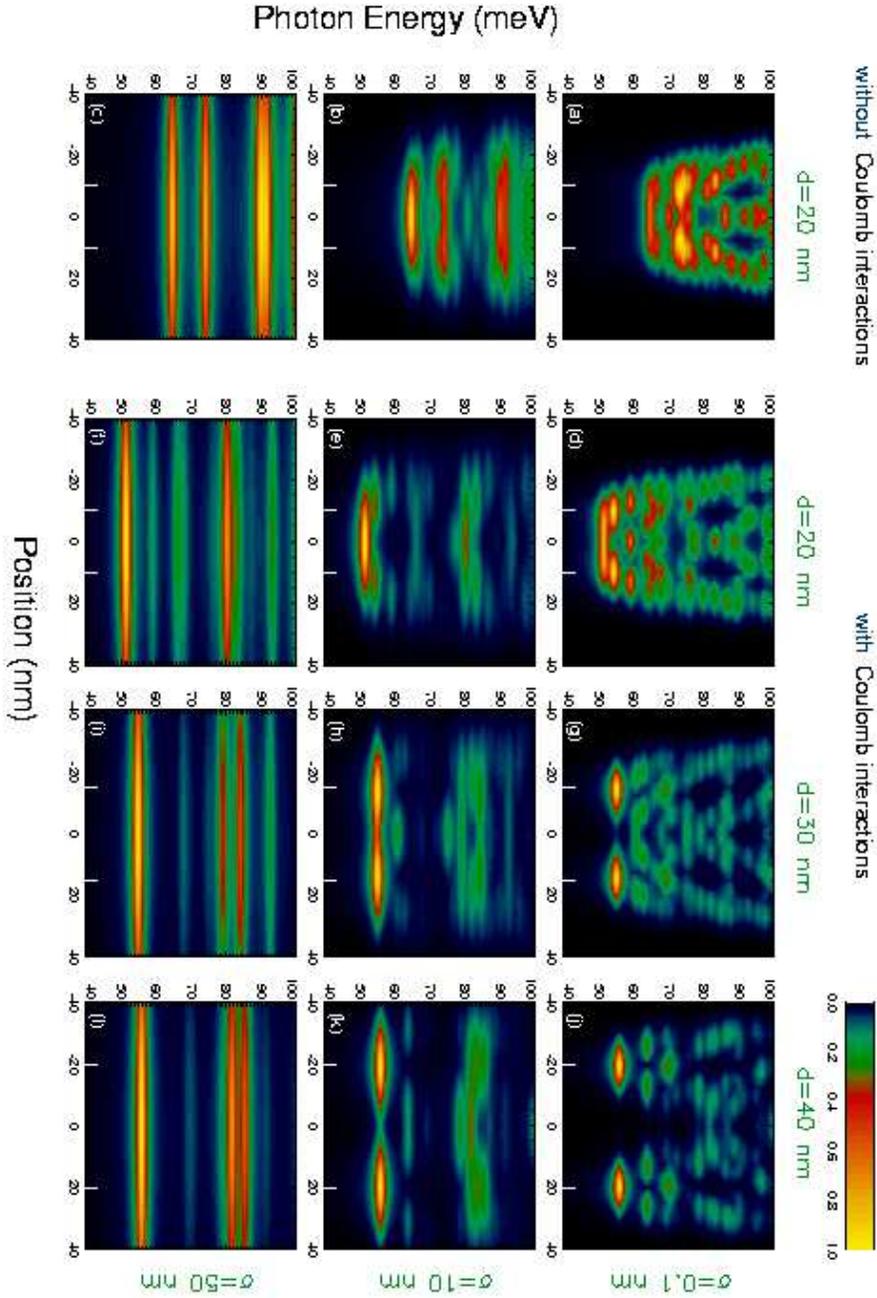,height=20cm}
\vspace*{0.7cm}
\caption
{Local absorption spectra $\alpha_\xi(X,\hbar \omega)$  for a double QD
with (Figs.~7(d--l)) and without (Figs.~7(a--c)) Coulomb interactions
and for different values of $\sigma$
and interdot distance $d$. Photon energy $\hbar\omega$ is
measured with respect to the bandgap, and $X$ is the position of the
tip along the $x$-axis ($Y=0$). In our calculations we use a basis of 12
electron and hole states, respectively.}
\end{figure}
\pagebreak

\begin{figure}
\vspace*{0.0cm}
\hspace*{0.4cm}\psfig{file=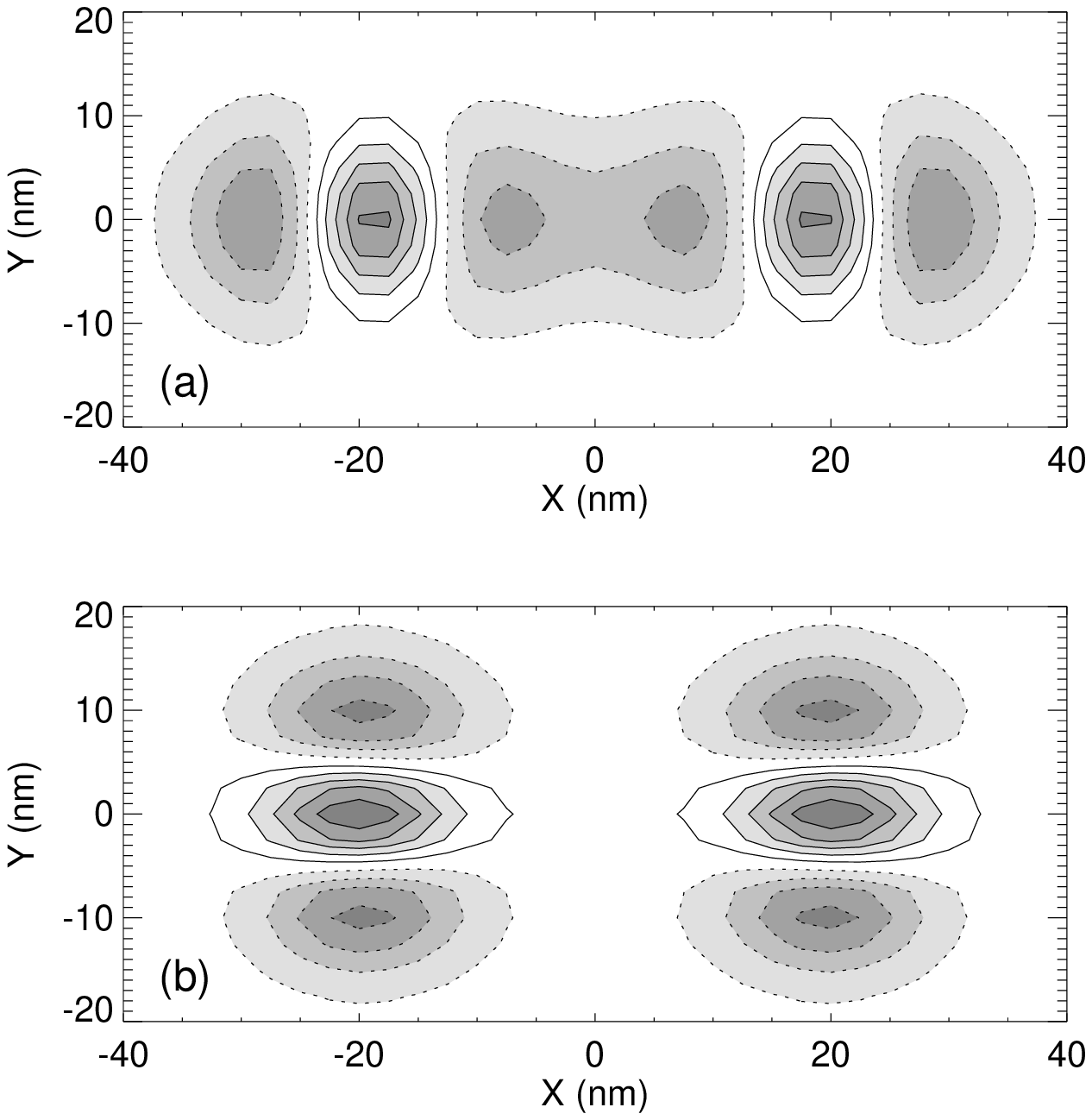,height=16cm}
\vspace*{0.2cm}
\caption
{Contour plot of the exciton wavefunction $\Psi^\lambda({\bf r},{\bf r})$ 
of the two excitons which are responsible for the non-monotonic behaviour of
the features at $\approx 70$ meV at the interdot distance $d=40$ nm.
Solid and dashed lines correspond to positive and negative values,
respectively. The upper (lower) panel refers to exciton with energy
69.1 meV (70.3 meV).}
\end{figure}
\pagebreak


\begin{thebibliography}{10}
\bibitem{basis} 
In our calculations, we use a basis of 12 electron and hole states,
respectively (corresponding to a total number of 144 exciton states).
\end{thebibliography}
\end{document}